\documentclass[a4paper,11pt]{article}
\usepackage{pos}
\usepackage{subcaption}
\usepackage{wrapfig}
\usepackage{gensymb}
\newcommand{\refer}[1]{Ref.~\cite{#1}}

\title{Enhancing air-shower observations: Results from an IceCube Surface Array Prototype Station}

\manuallySeparateAuthors
\author*[a]{Megha Venugopal} 
\author [b] {for the IceCube Collaboration}
\renewcommand{\printHeadAuthors}{}
\affiliation[a]{Institute for Astroparticle Physics, Karlsruhe Institute of Technology (KIT),\\
  Hermann-von-Helmholtz Platz 1, Eggenstein-Leopoldshafen, Germany}
\affiliation[b]{Full authorlist at \href{https://icecube.wisc.edu/collaboration/authors/\#collab=IceCube&date=2024-08-18&formatting=web}  
{https://icecube.wisc.edu/collaboration/authors/}
}
\emailAdd{megha.venugopal@kit.edu}

\abstract{The Surface Array Enhancement of the IceCube Neutrino Observatory is set to equip the existing surface cosmic-ray array of ice-Cherenkov detectors, IceTop, with radio antennas and scintillation detectors. This can lower the energy threshold of detection and increase the measurement accuracy of IceTop. The antenna readout uses a multiplicity trigger from the scintillation detectors. A fully functioning prototype station of the enhancement was deployed at the South Pole in 2020 and upgraded in 2023, and provides a pathway for the future IceCube-Gen2 surface array detector. Previous results include air-shower searches with existing data using traditional and machine learning methods and a preliminary estimation of $X_\mathrm{max}$. The detection methods for radio involving multi-detector components are described and the latest results are presented.}

\FullConference{10th International Workshop on Acoustic and Radio EeV Neutrino Detection Activities (ARENA2024)\\
11-14 June 2024\\
The Kavli Institute for Cosmological Physics, Chicago, IL, USA\\}

\begin{document}
\maketitle

\section{Introduction}
The IceCube Neutrino Observatory comprises a deep in-ice detector consisting of Digital Optical Modules (DOMs) with photo-multiplier tubes and the IceTop surface detector. One of the main objectives of IceCube is to detect astrophysical neutrinos. It also detects atmospheric neutrinos and muons. The DOMs capture Cherenkov light produced by interactions of these particles with ice resulting in secondary particles. IceTop, built with 162 ice-Cherenkov tanks, vetoes atmospheric neutrinos produced by cosmic-ray showers for the neutrino observatory. It has contributed to the understanding of the cosmic-ray spectrum in the PeV to EeV energy range \cite{S125IceTop}. Continuous winds at the South Pole cause snow accumulation on the detectors. The snow accumulation and resulting signal attenuation is not uniform across each detector on IceTop. This has led to a continuous decrease in efficiency in detecting lower-energy cosmic rays over the years as well as higher uncertainty in air-shower reconstruction. To overcome this, elevated scintillation detectors and antennas surrounding the existing IceTop detectors were proposed \cite{HaungsEnhancement}. The scintillation detectors can lower the measurement threshold and the antennas have a higher sensitivity to $X_\mathrm{max}$, the shower maximum \cite{FrankRadioPaper}. Together, an overall improvement in measurement accuracy can be expected. A prototype station with three antennas and eight scintillation detectors was deployed in 2020 and the data acquisition was improved in 2022. The detectors were also upgraded in 2023 \cite{ShefaliScintICRC}. The layout of the station can be found in Fig. \ref{fig:LayoutandAntenna}. 

Test stations with the same single station setup were also deployed at the Pierre Auger Observatory, Malargüe and the Telescope Array site, Utah. Initial results of the station at Malargüe can be found in \refer{StefARENA2024}.

\section{Instrumentation and Radio Signal Processing in the Enhancement}

\begin{figure}

\begin{subfigure}{0.6\textwidth}

\includegraphics[width=0.6\linewidth, height=8cm]{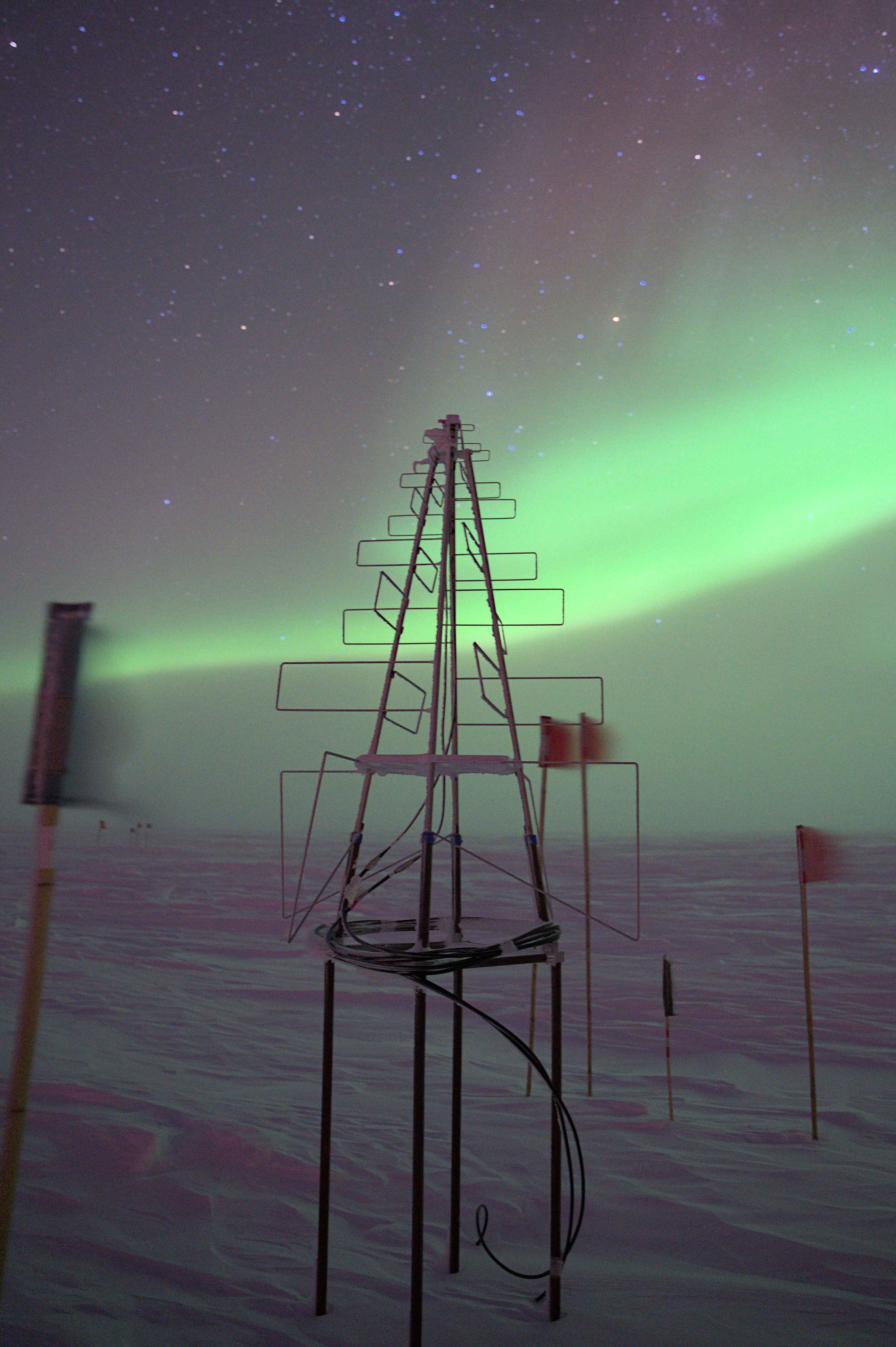}
\label{fig:SKALAAntenna}
\end{subfigure}
\hspace{-15mm}
\begin{subfigure}{0.5\textwidth}
\includegraphics[width=1\linewidth, height=6cm]{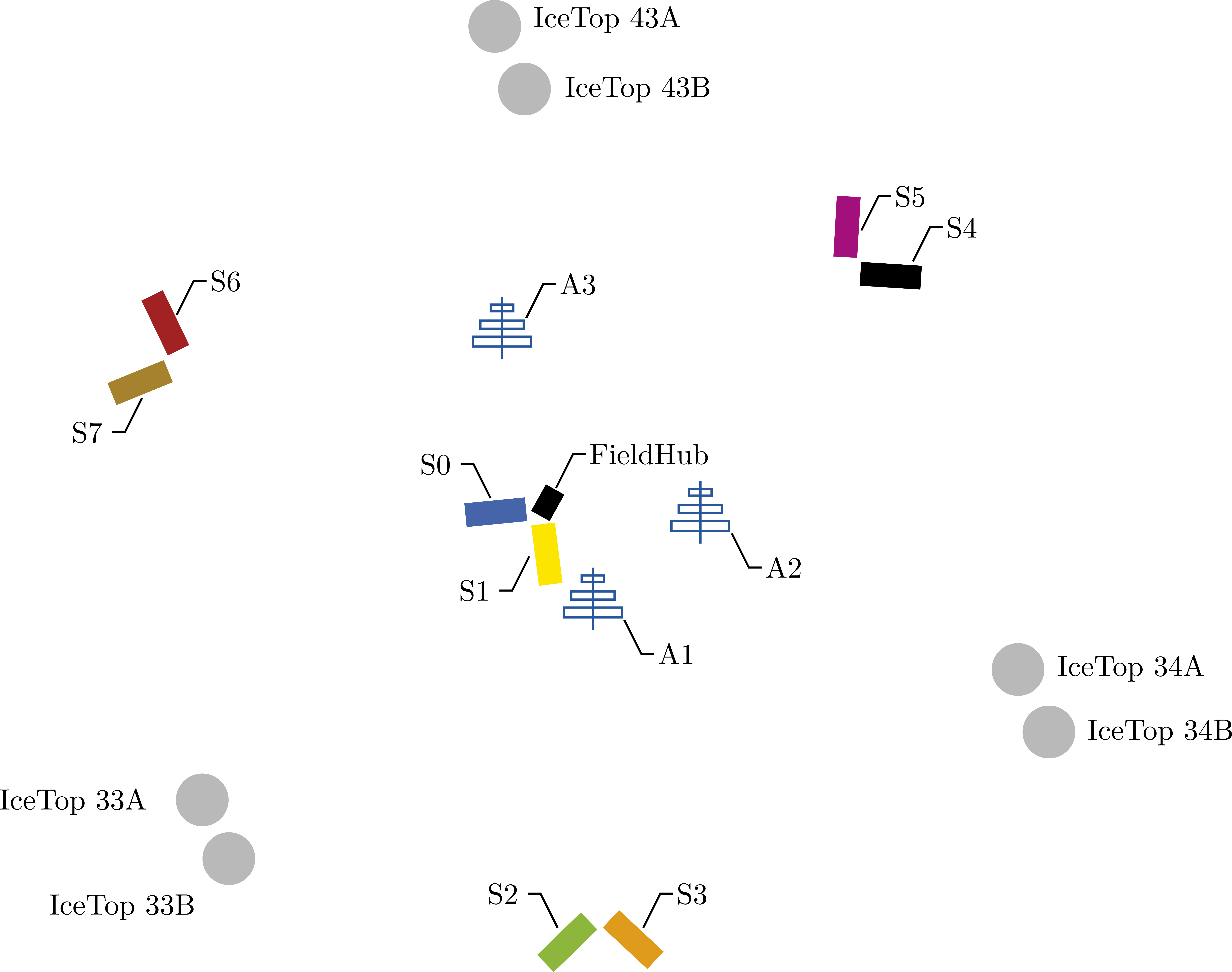} 
\label{fig:DetectorLayout}
\end{subfigure}
\caption{Left: A single SKALA v2 antenna with the aurora australis in the background. Right: The layout of the Prototype station currently deployed at the South Pole with the nearby IceTop stations. 
}
\label{fig:LayoutandAntenna}
\end{figure}

The hybrid data acquisition system of the enhancement of IceTop powers and controls both the scintillation detectors and the antennas. The scintillation detectors capture the light emitted by ionizing particles in scintillation bars, which is collected by optical fibers and routed to a Silicon photomultiplier. The signal is then sent to a board (termed uDAQ) where the signal is read out and digitized. A detailed overview and the current status of the scintillation detectors can be found in \refer{ShefaliScintICRC}. The antennas used are the second generation (see image in Fig. \ref{fig:LayoutandAntenna}) of the ones proposed for the SKA experiment \cite{SKALAV2}, which for the enhancement can operate in the frequency band from 70 -- 350\,MHz. 

The antennas measure mutually perpendicular components of the radio-signal which are treated independently both in hardware and in analysis. The signal passes through a two-board Low Noise Amplifier (LNA) where it is amplified by 40\,dB while reducing any additional noise from the board. It is powered by a bias-tee that is on a radio board, termed radioTAD, which also filters the signal into a specified measurement range with a high pass and low pass filter. The signal is then split into 4 parts per polarization to be sent to the DRS4 ring buffer with (8+1) channels \cite{thesisRox}.  
\begin{figure}
     \centering
\includegraphics[width=0.6\linewidth]{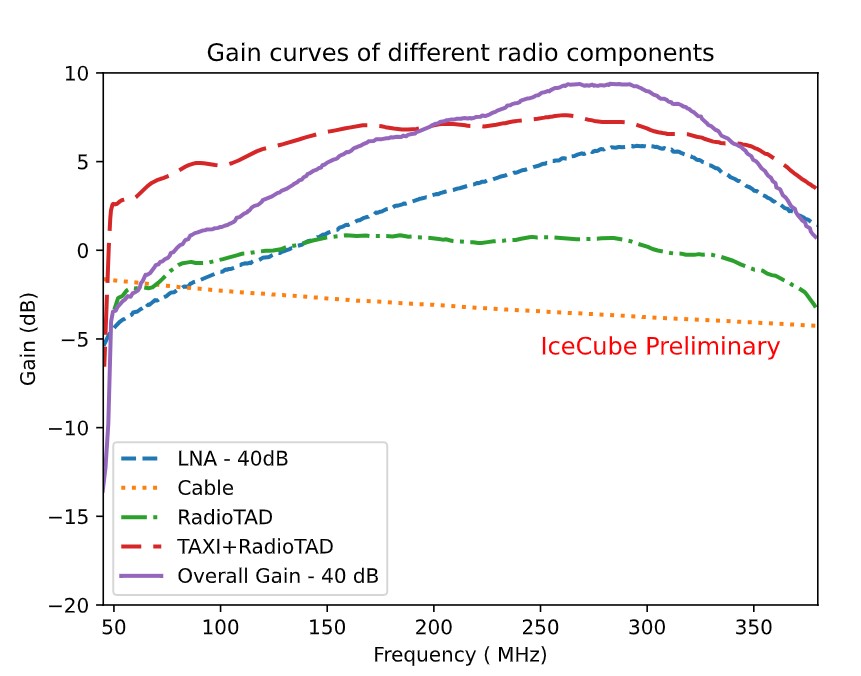} 
    
     \caption{Gain of different components in the radio signal chain. For both the LNA and the overall gain, the values have been reduced by 40 dB for representation purposes.}
     \label{fig:GainCurves}
     
 \end{figure}
 
\section{Data acquisition and calibration}

The trigger from the scintillation detectors starts the readout of the antennas. A multiplicity condition of 6 or more above threshold signal-hits in a 1\,\textmu s window is used. Until the trigger is sent, the data is continuously stored in the DRS4 ring buffer. Once the signal is read out, it is sampled and then digitized with an ADC with a sampling rate of 1\,GHz. 
The data acquisition is housed in a shielded box and the whole setup is referred to as the TAXI \cite{TAXI}. Timing information is provided with a White Rabbit (WR) node where the central timing information from the IceCube Laboratory is synchronized with the board in the TAXI. The TAXI is currently independent of the IceCube and IceTop electronics. 
Different components are calibrated in the lab with different methods, some presented in \cite{thesisRox} and the gain curves are shown in Fig.\ref{fig:GainCurves}.

The measurement of the gain for the TAXI was taken with the entire chain to include readout response. We notice a relatively smooth overall gain in our measurement range. 

\section{Method of air-shower identification}
The identification of air showers was done using two methods. One used a traditional Signal-to-Noise Ratio (SNR) study that is discussed in this work and the other one with a higher rate of identifying events used a machine learning method, details of which can be found in \refer{AbdulFrankARENA2024}. The data that was considered for this work was taken from January to July 2022 where certain periods were not taken into account due to differences in data taking. 
Although the radio buffer readout is capable of storing 4\,\textmu s of data, we only use the setting to readout 1\,\textmu s of data. This was assessed to be sufficient to identify even highly inclined showers within the footprint of the prototype station. It also improves the data quality due to the fourfold redundancy of each sample that is stored in the ring buffer. A coincident event is created in the data processing by combining information from the radio antennas, scintillation detector and IceTop. This is done by checking if the timestamps of these events recorded in each detector lie within a 2\,\textmu s window.

The median of the four redundant readout channels of the DRS4 is taken for each polarization of each antenna. Any identified readout artifact is cleaned. The pedestal is removed and the baseline of the signal is brought back to zero before converting the signal from ADC units to voltage. 
The spike filter uses the median background spectra and suppresses those frequency bins with higher noise from the event traces. This method of cleaning Radio Frequency Interference (RFI) is further discussed in \cite{ColemanFreqFilter}. The data is filtered into the band of 100 -- 230\,MHz and then the electronics response is removed. This includes the custom response of individual components including the LNA, cable and the DAQ. In order to calculate the threshold for the SNR cut, background antenna data recorded using a fixed rate trigger is utilized.
The SNR is computed by dividing the time series into equal chunks and computing the squared peak in each trace to the median RMS of the traces in each chunk. Hence,
\begin{equation} 
\label{eq:SNR}
SNR = \left(\frac{\text{Peak Amplitude in each trace}}{
\text{Median of RMS chunks}}\right)^2 .
\end{equation}

A threshold for a minimum SNR is applied such that 95\% of background rejection in each antenna channel is performed using this method. The SNR of each antenna is computed independently and air-shower events are identified in scintillator triggered events by requiring that each antenna has at least one channel above threshold.
Directional reconstruction is performed with the data from the three antennas using the expectation from a plane wavefront fit. 

\section{Directional Reconstruction}
\begin{figure}
\begin{subfigure}{0.5\textwidth}
\includegraphics[width=1\linewidth, height=5cm]{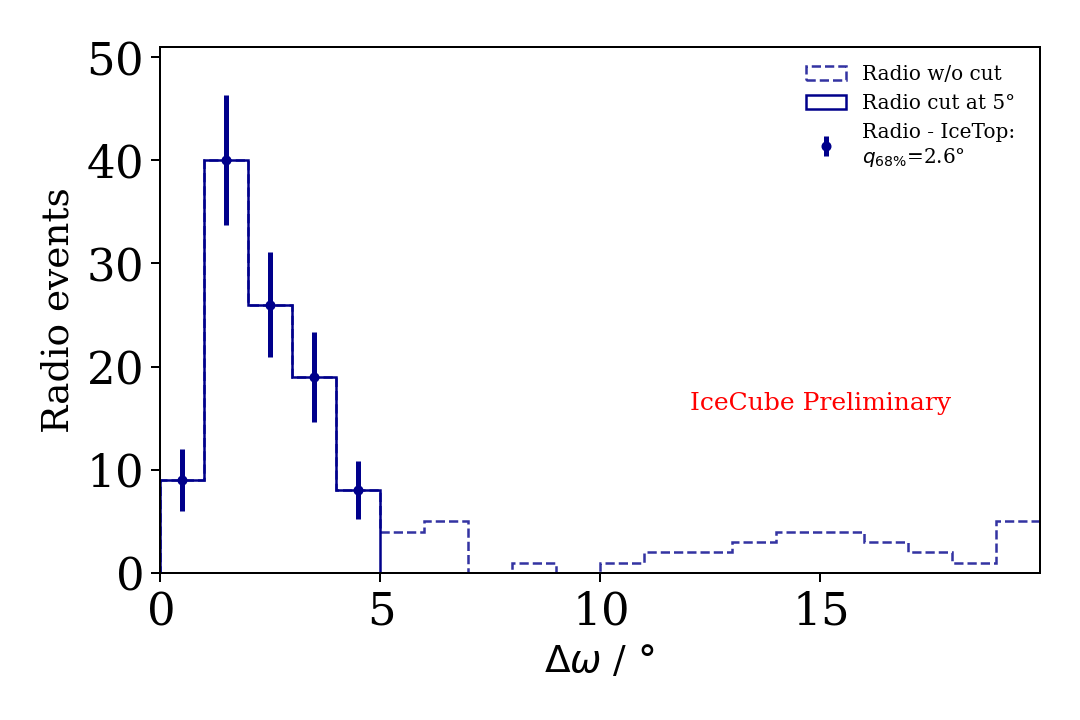} 
\label{fig:OmegaCut}
\end{subfigure}
\begin{subfigure}{0.6\textwidth}
\hspace{-5mm}
\includegraphics[width=1\linewidth, height=6cm]{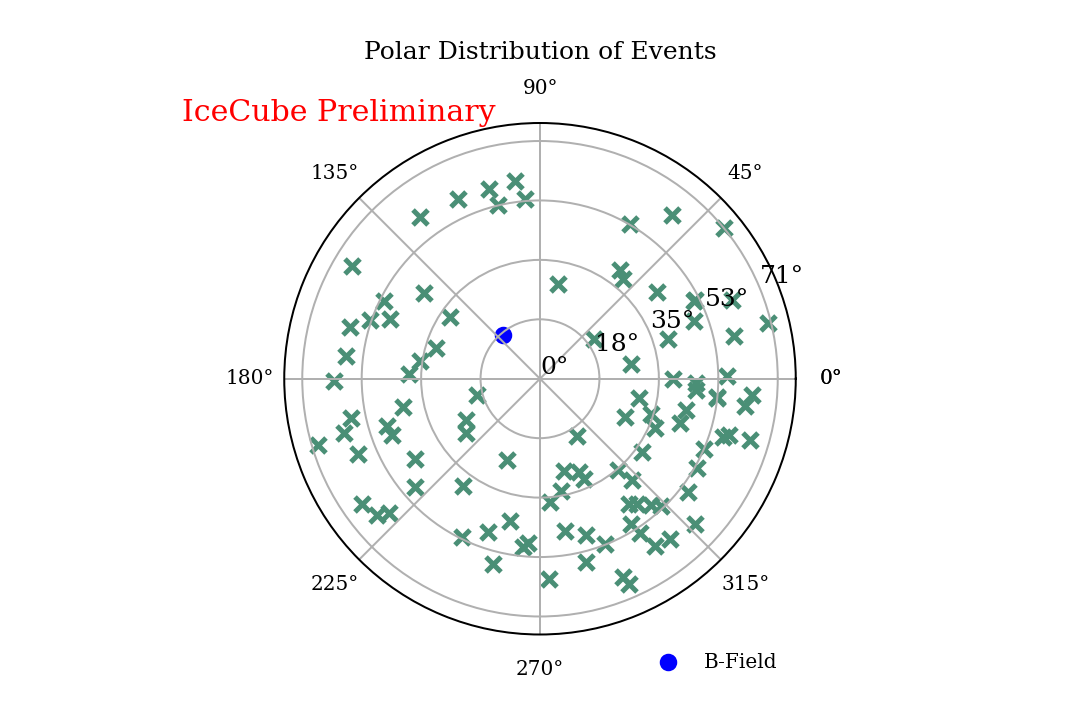}
\label{fig:PolarPlot}
\end{subfigure}
\caption{Left: The figure shows the distribution of opening angles with respect to the IceTop reconstructed direction for all air shower events identified with radio antennas. The dashed lines indicate the events thrown away after the 5\degree\:cut. Right: The sky distribution of these events. The magnetic field direction at the South Pole is indicated in blue.}
\label{fig:OpeningangleandPolarPlot}
\end{figure}

\begin{figure}
     \centering
     \includegraphics[width=0.6\linewidth]{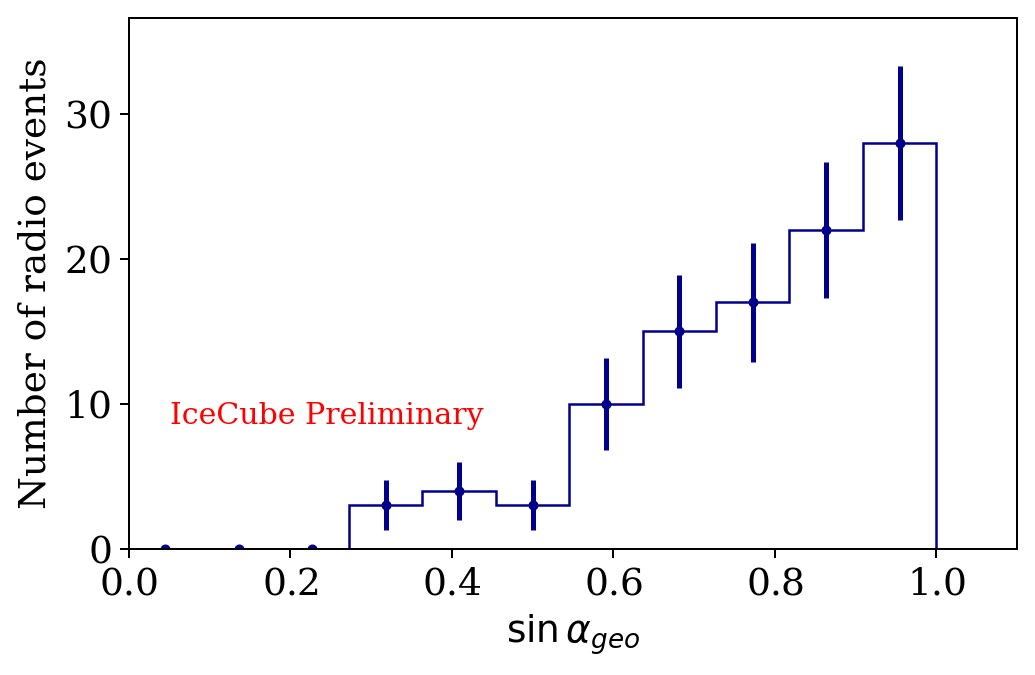}
     \caption{Histogram of identified cosmic-ray showers increasing with increasing geomagnetic angle}
     \label{fig:GeomagneticScaling}
\end{figure}

The reconstructed data is compared to the existing IceTop detector. An additional cut is made to reject events with an opening angle greater than 5\degree\: with respect to IceTop reconstruction, as shown in Fig.~\ref{fig:OpeningangleandPolarPlot}. After the cut, 102 events are left for analysis. The  distribution of these events in the sky is presented in Fig.~\ref{fig:OpeningangleandPolarPlot}. There are fewer events in the sky around the direction of the geomagnetic field at the South Pole.
It is well known that the radio emission from cosmic rays scales with the sine of the geomagnetic angle \cite{TimRadioDetection,FrankRadioPaper}, which is the angle between the primary cosmic ray and the geomagnetic field. The identified events are shown as a function of the sine of the geomagnetic angle in Fig.~\ref{fig:GeomagneticScaling}. We observe an increase with increasing geomagnetic angle as expected. This further confirms the veracity of the identified events. 

\begin{figure}
     \centering
     \includegraphics[width=0.6\linewidth]{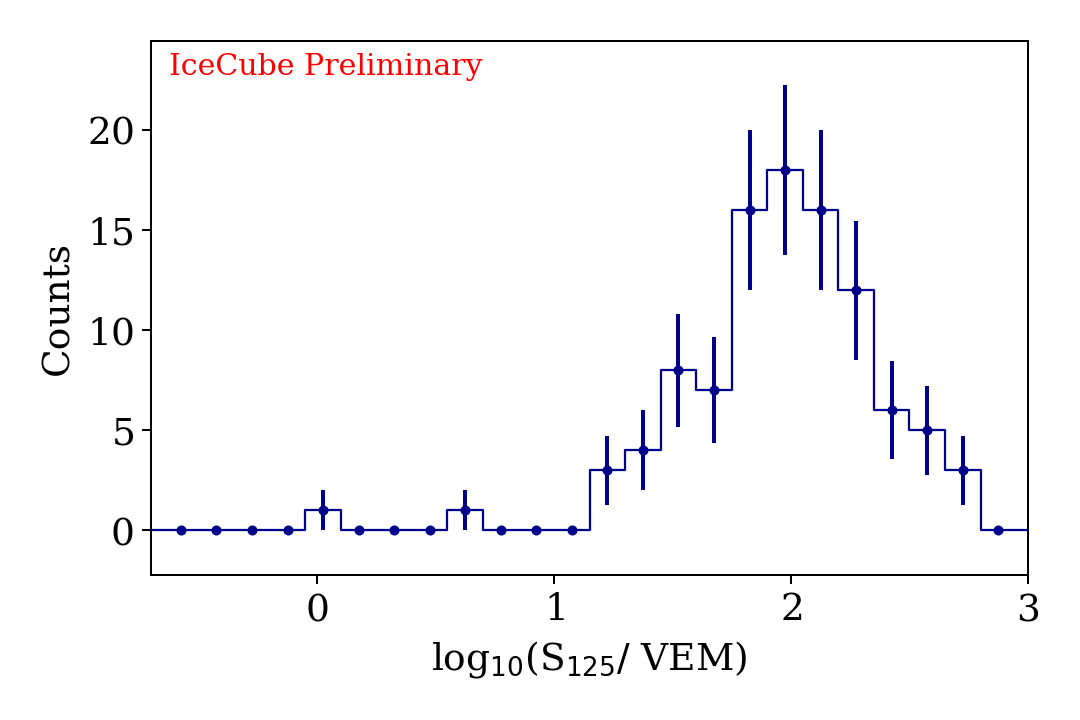}
         \caption{ $\log_{10} (S_{125}/VEM)$ distribution of identified air-shower events normalized to a Vertical Equivalent Muon (VEM). The lower energy events can be explained by false positive events expected due to background passing through the air-shower identification pipeline. We expect approximately two events to be false positives.}
     \label{fig:S125dist}
\end{figure}

\begin{figure}

\begin{subfigure}{0.5\textwidth}
\centering
\includegraphics[width=1\linewidth, height=6cm]{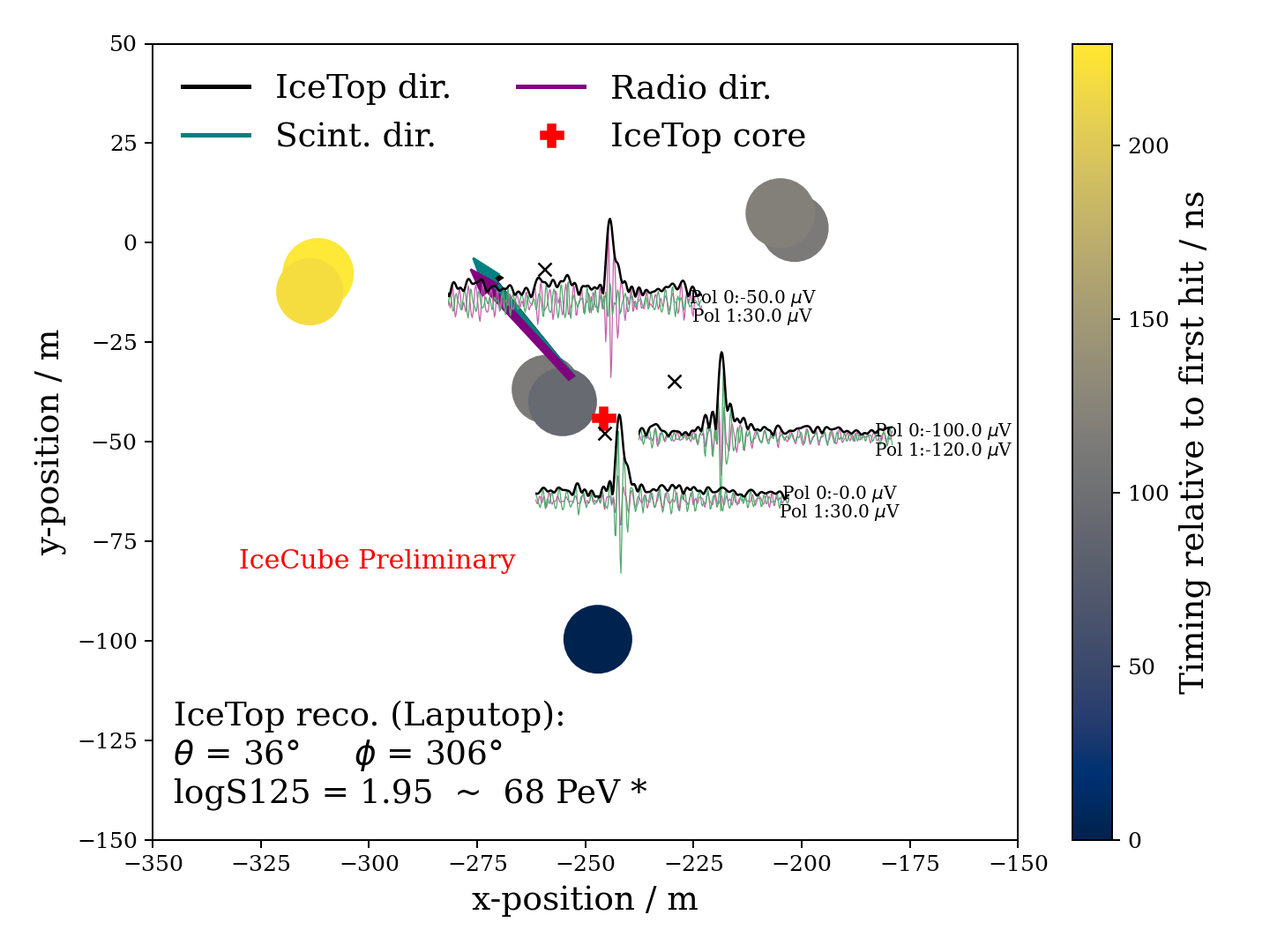} 
\label{fig:Corein}
\end{subfigure}
\hfill
\begin{subfigure}{0.5\textwidth}
\centering
\includegraphics[width=1\linewidth, height=6cm]{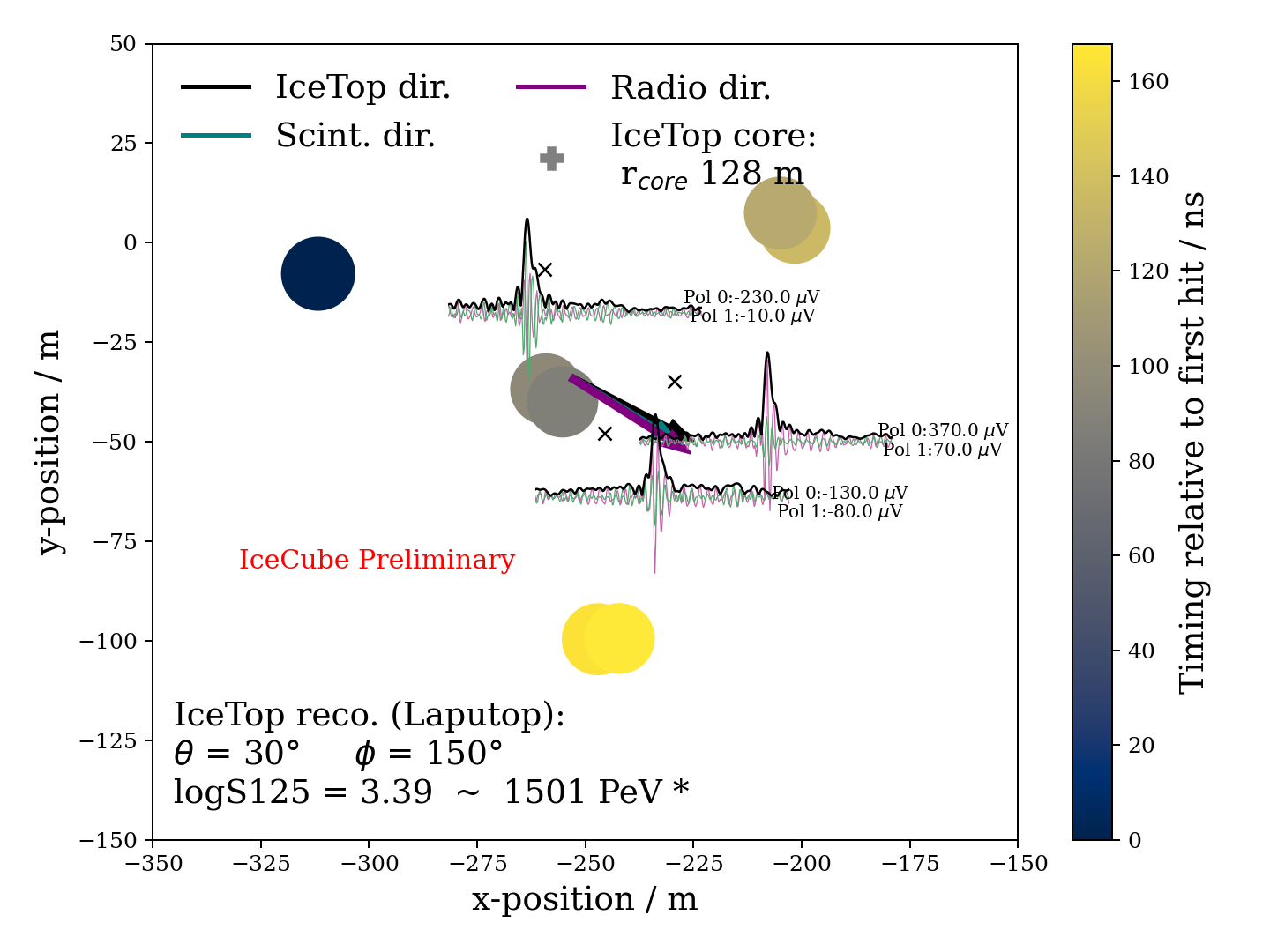}
\label{fig:Coreout}
\end{subfigure}
\caption{The figure shows two events measured in three different detectors: radio antennas, scintillation detectors and IceTop. The location of the scintillation detector is represented with circular blobs. The scintillation detector charge is represented by  size of the blobs while the colour shows the timing of hit with respect to the first hit. The arrows indicate reconstructed directions with different detectors. The red plus sign indicates core reconstructed within the figure and the grey plus sign for the core lying outside. The crosses represent the positions of the antenna and the corresponding pulses recorded along with the Hilbert envelope.} 
\label{fig:Threefold}
\end{figure}

The quantity $S_\mathrm{125}$ is the energy proxy for IceTop providing a reliable calibration for showers with a zenith angle lower than 36.8\degree. It is defined as the expected signal strength at a reference distance of 125\,m from the shower axis. For inclined showers, it does not give an accurate estimate of energy. The $S_\mathrm{125}$ distribution of the identified events is given in Fig. \ref{fig:S125dist}. False positive events are calculated by passing pure background data through the processing pipeline. The air showers identified are reconstructed after which the opening angle cut of 5\degree\:is made. Two false positive events are expected and in Fig. \ref{fig:S125dist} we see two events with a $\log_{10} (S_{125}/VEM)$ less than 1. These can be potentially explained as the false positive events.
 Example events as measured by the scintillation detector, radio antennas and the IceTop detectors are presented in Fig. \ref{fig:Threefold} to show the threefold coincidence.

\section{Conclusion}
Air-shower events identified with the radio antennas of a prototype station at the South Pole have been presented and a coincidence study has been conducted with respect to IceTop. The multi-component detector system offers a reliable means of identifying air showers. During the prototype stage, it helps to cross-check the reconstruction parameters with respect to the existing IceTop detector. Over the course of the measurement period we have detected over 100 events. Example events measured in all three detectors are also shown.

With the deployment of further stations, a more robust and independent radio reconstruction can be expected. The successful air-shower detection also demonstrates the readiness of the station design for a larger array enhancing IceTop and for the future IceCube-Gen2 Surface Array~\cite{IceCube-Gen2Whitepaper}.
\bigskip
\bigskip
\bigskip
\bigskip
\bigskip
\bigskip
\bigskip
\bigskip
\bigskip
\bigskip
\bigskip
\bigskip
\bibliographystyle{ICRC}
\bibliography{refs}
\section*{Acknowledgement}
The author would like to acknowledge the support of the German Academic Exchange Service (DAAD) and Karlsruhe School of Elementary Particle and Astroparticle Physics: Science and Technology (KSETA) for funding and making this research possible. 


\end{document}